\documentclass[aps,prl,twocolumn,groupedaddress,superscriptaddress,showpacs]{revtex4}

\usepackage{amsmath}
\usepackage{graphicx}
\usepackage{bm}
\usepackage{hyperref}

\newcommand{\Hca}{\mathcal{H}}

\newcommand{\tp}{t_{\perp}}
\newcommand{\kpa}{k_{||}}
\newcommand{\kpe}{k_{\perp}}

\newcommand{\SA}{\Sigma_A}
\newcommand{\SB}{\Sigma_B}
\newcommand{\SAo}{\Sigma_A (\omega)}
\newcommand{\SBo}{\Sigma_B (\omega)}

\newcommand{\GR}{{\rm G}}
\newcommand{\GRb}{\overline{\rm G}}

\newcommand{\GDA}{{\rm G}_{\rm AA}^{\rm D}}
\newcommand{\GDB}{{\rm G}_{\rm BB}^{\rm D}}

\newcommand{\GDAb}{\overline{\rm G}_{\rm AA}^{\rm D}}
\newcommand{\GDBb}{\overline{\rm G}_{\rm BB}^{\rm D}}

\newcommand{\nimp}{n_{\rm i}}

\newcommand{\e}{\epsilon}

\newcommand{\om}{\omega}

\newcommand{\vk}{{\bf k}}

\newcommand{\vf}{v_{\rm F}}

\newcommand{\nf}{n_{\rm F}}

\newcommand{\Ima}{{\rm Im}}
\newcommand{\Rea}{{\rm Re}}

\newcommand{\Hkin}{\mathcal{H}_{0}}

\begin{document}

\title{Electronic properties of graphene multilayers}

\author{Johan Nilsson}
\affiliation{Department of Physics, Boston University, 590 
Commonwealth Avenue, Boston, MA 02215, USA}

\author{A.~H. Castro Neto}
\affiliation{Department of Physics, Boston University, 590 
Commonwealth Avenue, Boston, MA 02215, USA}

\author{F. Guinea}
\affiliation{Instituto de  Ciencia de Materiales de Madrid, CSIC,
 Cantoblanco E28049 Madrid, Spain}

\author{N.~M.~R. Peres}
\affiliation{Center of Physics and Departamento de F{\'\i}sica,
Universidade do Minho, P-4710-057, Braga, Portugal}

\date{\today}

\begin{abstract}
We study the effects of disorder in the electronic
properties of graphene multilayers, with special focus on the
bilayer and the infinite stack. At low energies and long
wavelengths, the electronic self-energies and 
density of states exhibit behavior with divergences near half-filling. 
As a consequence, the spectral functions and conductivities 
do not follow Landau's Fermi liquid theory. In particular, we
show that the quasiparticle decay rate has a minimum as
a function of energy, there is a universal minimum value for the
in-plane conductivity of order $e^2/h$ per plane and, unexpectedly, 
the c-axis conductivity is enhanced by disorder at low doping,
leading to an enormous conductivity anisotropy at low temperatures.
\end{abstract}


\maketitle

{\it Introduction.} Recently, many properties of a single graphene sheet 
have been studied theoretically by several groups. These properties are, in many cases, found to be unconventional due to the peculiar band structure of graphene which is described in terms of Dirac fermions at the edge of the Brillouin zone (BZ). This activity was triggered by the realization that graphene could be obtained and studied experimentally \cite{Netal04_short}, and the subsequent experiments that followed to further characterize the material \cite{Novolelov2005_short,Zhang2005_short}.

More recently the attention has turned to graphene multilayers \cite{berger04_short} and, in particular, to bilayers that also show 
anomalies in the integer quantum Hall effect (IQHE) \cite{Novoselov2006_bilayer_short} and have received theoretical attention \cite{Falko2006,Nilsson2005exchange_short}. 
In this paper we show that the bilayer graphene has also very 
unconventional behavior in its spectral and transport properties.
These properties, however, are quite different from those of the
single layer. In fact, the anomalous behavior is a property
of all multilayer graphene systems we have studied. 
It is also worth noting that the higher complexity of the multilayer
systems may be of interest for device applications since it also 
allows for greater flexibility in tailoring the electronic properties.

There are two key ingredients that make the physics of the graphene 
multilayers unconventional. Firstly, due to the relatively weak 
interlayer coupling it inherits some properties from its parent 
material, the single graphene sheet. The existence of Dirac points 
in the BZ, where the electron and hole bands becomes 
degenerate, arises from the physics of the single layer in conjunction 
with the  second important ingredient that is the peculiar geometry that
results from the A-B stacking of the planes. This geometry implies that 
the binding of the planes due to orbital overlap is mainly sitting on 
one of the sublattices (that we call A, the other sublattice being B) 
in each plane (the different planes of the units cell are denoted by 
1 and 2). The main residence of the low-energy states is then on the
sublattice B. Nevertheless, electron transport coming from nearest 
neighbor hopping must go over the A sublattice. This feature implies 
that even though the total density of states at half-filling is finite, 
the density of states on the A sublattice vanishes as the Dirac point 
is approached, leading to unconventional in-plane and out-of-plane 
transport properties. The main purpose of this work is to show how 
this comes about and to highlight the unusual behavior of the 
self-energies due to disorder near half-filling. 

{\it Graphene bilayer}. 
At low energies and long wavelengths, one can expand the electronic 
spectrum close to the K and K' points in the BZ, leading
to a Hamiltonian of the form \cite{Wallace47}: $H = \sum_{\vk}
\Psi^{\dag}(\vk) \Hkin(\vk) \Psi(\vk)$, where ${\vk}=(k_x,k_y)$ is 
the two-dimensional momentum measured relative to the $K$ ($K'$) 
point, 
\begin{equation}
  \label{eq:Hkin0bilayer}
  \Hkin(\vk) = \vf
  \begin{pmatrix}
    0 & k e^{i \phi(\vk)} & \tp / \vf & 0 \\
    k e^{-i \phi(\vk)} & 0 & 0 & 0 \\
    \tp / \vf  & 0 & 0 & k e^{-i \phi(\vk)} \\
    0 & 0 & k e^{i \phi(\vk)} & 0
  \end{pmatrix},
\end{equation}
$\Psi^{\dag}(\vk) = \bigl( c^{\dag}_{\text{A}_1,\vk} \; c^{\dag}_{\text{B}_1,\vk} \;
c^{\dag}_{A_2,\vk} \; c^{\dag}_{\text{B}_2,\vk} \bigr)$ is the electron 
spinor creation operator, $\phi(\vk) =
\tan^{-1}(k_{y}/k_{x})$ is the two-dimensional angle in momentum space, 
$\vf = 3 t a / 2$ is the Fermi-Dirac velocity, $t \approx 3 \, \text{eV}$ 
is the in-plane hopping energy, $a \approx 1.42 \, {\rm \AA}$  is the 
interatomic distance within the layers, $\tp \approx 0.35 \, \text{eV}$ 
is the interlayer hopping energy \cite{BCP88}. The $2\times2$-blocks on 
the diagonal of (\ref{eq:Hkin0bilayer}) are identical to the continuous 
approximation that leads to the massless Dirac spectrum in a single layer. 
In what follows we use units such that $\vf=1=\hbar$, and suppress the 
spin and valley indices unless otherwise specified. 
The Hamiltonian (\ref{eq:Hkin0bilayer}) can be easily diagonalized and 
the energy spectrum is made out of four bands with energy: 
$\tp/2 \pm E(k)$ and $-\tp/2 \pm E(k)$, where $E(k)=\sqrt{\tp^2/4 + k^2}$.
The resulting spectrum is made out of two vertex touching hyperbolae 
at zero energy, separated by a gap of energy $\tp$ from two other hyperbolae.

In what follows it is convenient to introduce a local frequency dependent 
self-energy which, due to the lattice structure, has different values in 
the A ($\SAo$) and the B ($\SBo$) sublattices. We take these into account 
by a diagonal matrix $\Hca_{\Sigma}(\om)$ and the Green's function is then 
given by:
\begin{equation}
  \label{eq:green1}
  \text{G}^{-1}(\om,\vk) = \om - \Hkin(\vk) - \Hca_{\Sigma}(\om).
\end{equation}
Because the Hamiltonian factorizes into two blocks it is simple to work 
out the explicit expression for the Green's function. G$(\om,\vk)$ is a 
$4 \times 4$ matrix, but for our purposes here the most important components 
are the diagonal ones that are given by:
\begin{eqnarray}
\GDA(\om,k) &=& \sum_{\alpha=\pm}(\om - \SB(\om))/ (2 D_{\alpha}(\om,k)) \, ,
\nonumber \\
\GDB(\om,k) &=&  \sum_{\alpha=\pm}(\om - \SA(\om) + \alpha \tp) / (2 D_{\alpha}(\om,k)) \, ,
\nonumber \\
D_{\pm}(\om,k) &=& \bigl[\om \pm \tp - \SAo \bigl] \bigl[\om -\SBo \bigr] - k^2 \, .
\nonumber
\end{eqnarray}

{\it Effects of disorder.} We use standard techniques to study the effect 
of disorder and average over impurity positions to get the disorder-averaged 
propagators \cite{nuno2006_short}. The effect of repeated scattering from a 
single impurity can be encoded in a self-energy which can be computed in the 
T-matrix approximation as:
\begin{eqnarray}
  \label{eq:sigma1}
  \Sigma(\om) &=& V \bigl[1 - \GRb(\om) V]^{-1}/N \, ,
\\
\GRb(\om) &=& \sum_{\vk} \GR(\om,\vk)/N \, ,
\nonumber 
\end{eqnarray}
where $V$ is the strength of the impurity potential, $N$ is the number of 
units cells in each plane. We turn the momentum sum into an integral by 
the introduction of a cut-off,  $\Lambda$ ($\approx 1/a$), that we estimate 
($\Lambda \approx 7 \, \text{eV}$) by a Debye approximation conserving 
the number of states in the BZ \cite{Nilsson2005exchange_short}. Due to the 
special form of the propagator and the impurity potential, the self-energy 
is diagonal. The result for unitary scattering (or site dilution) is obtained
by introducing a finite density of vacancies $\nimp$ and taking the limit 
$V \rightarrow \infty$, and one finds:
\begin{equation}
  \label{eq:GbarAA1}
  \SAo
  = -\nimp/\GDAb (\om)
\, ,
\quad
  \SBo
  = -\nimp/\GDBb (\om).
\end{equation}
In the full Born approximation (FBA) 
one uses the bare propagators on the right hand
side of (\ref{eq:GbarAA1}), the resulting self-energies are linear in the impurity
concentration so that this approximation neglects all 
correlations between different scattering centers. 
In the coherent potential approximation (CPA) \cite{Soven67} 
one assumes that the electrons are moving in an
effective medium that is characterized by the self-energies
$\SA$ and $\SB$. These must be determined
self-consistently by using the dressed propagators on the right hand
side of (\ref{eq:GbarAA1}), thus this approximation includes some 
effects of correlations between the scattering events (it
does not describe Anderson localization).
Using the explicit form of the propagators
we obtain the following self-consistent equations:
\begin{eqnarray}
  \frac{\nimp}{\Sigma_A}
  &=&
  \frac{\om - \SB}{2 \Lambda^2}
  \sum_{\alpha =\pm}
  \ln \Bigl[
  \frac{\Lambda^2}
  {- (\om + \alpha \tp -\SA) (\om -\SB)}
  \Bigr], 
  \nonumber \\
  \frac{\nimp}{\Sigma_B}
  &=&
  \frac{\om - \Sigma_A}{2 \Lambda^2}
  \sum_{\alpha =\pm}
  \ln \Bigl[
  \frac{\Lambda^2 }
  {- (\om -\alpha \tp -\Sigma_A) (\om -\Sigma_B)}
  \Bigr]
  \nonumber \\
  &+&
  \frac{\tp}{2 \Lambda^2}
  \ln \Bigl[
  \frac{- (\om -\tp -\SA) (\om -\SB)}
  {- (\om +\tp -\SA) (\om -\SB)}
  \Bigr].
\label{eq:CPAbilayer}
\end{eqnarray}
These equations include intervalley scattering in the intermediate states.
It is also straightforward to obtain the density of states (DOS) on sublattice
$a$ ($a=A,B$) from $\rho_{a}(\om) = -\Ima \overline{\text{G}}_{aa}^{\text{D}}
/ \pi$. In the clean case, one gets:
\begin{eqnarray}
\rho_{\text{A}}^{0}(\om) &=& |\om| \bigl[1+\Theta(|\om|- \tp ) \bigr] / 2
\Lambda^2 \, ,
\nonumber \\
\rho_{\text{B}}^{0}(\om) &=& \rho_{\text{A}}^{0}(\om) + \tp
\bigl[1-\Theta(|\om|- \tp) \bigr] / 2 \Lambda^2 \, ,
\nonumber
\end{eqnarray}
where $\Theta(x)$ is the Heaviside step function. Notice that the DOS 
on the A sublattice vanishes at zero energy (as in the case of a single 
layer), while $\rho_{\text{B}}^{0}(0)$ is finite.

\begin{figure}[htb]
\includegraphics[scale=0.43]{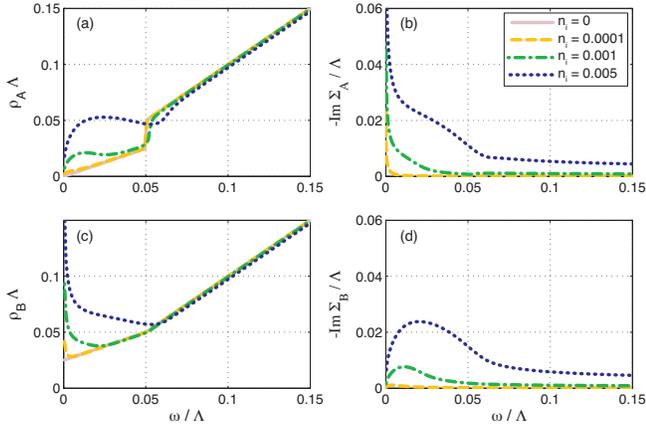}
\caption{(Color online) Bilayer DOS and self-energy (in units of the $\Lambda$)
in the CPA approximation. DOS on the A{\bf (a)} and B{\bf (c)} sublattice as a 
function of the frequency (in units of $\Lambda$), imaginary part of the 
self-energy on the A{\bf (b)} and B{\bf (d)} sublattice as a function of 
the frequency.}
\label{fig_selfenergy_dos_bilayer}
\end{figure}

We have solved the CPA equations in Eq. (\ref{eq:CPAbilayer}) and an 
example of  the resulting self-energies and the corresponding DOS are shown 
in Fig. \ref{fig_selfenergy_dos_bilayer}. Within the FBA $\SA$ diverges as 
$\nimp \Lambda^2/ \om$ up to logarithmic corrections as $\om \rightarrow 0$. 
In the single layer, the CPA makes the self-energy finite at $\om =0$. 
In contrast, the bilayer (and the multilayer) does not allow a finite 
$\SA$ at $\om =0$ in the CPA. One can see this by studying 
Eq.~(\ref{eq:CPAbilayer}) at $\om = 0$. Then the last line of
Eq.~(\ref{eq:CPAbilayer}) divided by $\SA$ must vanish, 
and this is not possible for finite values of $\SA$, furthermore
$\SB$ must vanish in this limit. These results imply that,
contrary to the single plane case, $\rho_{\text{A}}(\om \rightarrow 0)$ is 
zero even within the CPA. More explicitly, by defining 
$\SA \SB = - \xi \Lambda^2$ one can show (assuming $\SA \gg \tp$ and 
$\SB \gg \om$) that $\SA$ and $\SB$ are given asymptotically by:
\begin{subequations}
  \label{eq:S_smallOm}
\begin{eqnarray}
  \SA(\om) &\approx&e^{-i \pi/3} \left( \tp^2 \xi^2 \Lambda^2/\nimp\right)^{1/3} \om^{-1/3}  
  \\
  \SB(\om) &\approx& e^{-i 2 \pi/3} \left(\nimp \Lambda^4 \xi/\tp^2 \right)^{1/3} \om^{1/3} ,
\end{eqnarray}
\end{subequations}
for $\om \ll \Lambda^2 \xi^2 / \nimp \tp$, where $\xi$ is given by $\nimp =
\xi \ln(1 / \xi)$. Hence, the self-energies are rather anomalous and clearly
violate Landau's Fermi liquid theory. Note that 
$\sqrt{\xi} \Lambda \sim \sqrt{\nimp} \Lambda$ is exactly the
scale generated by disorder in the single plane case \cite{nuno2006_short}.

\begin{figure}[htb]
\includegraphics[scale=0.40]{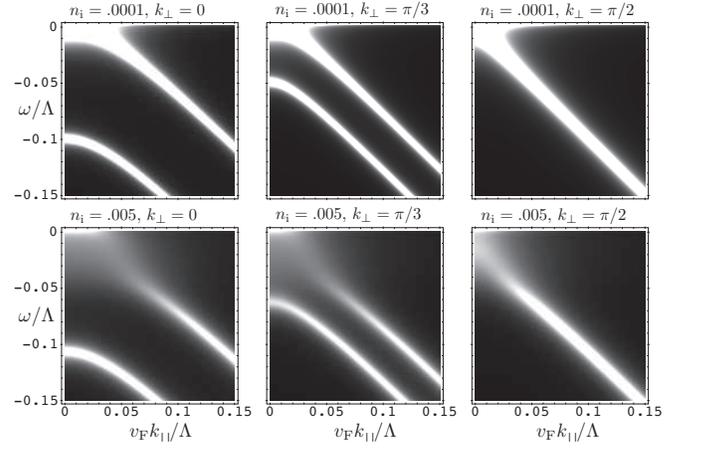}
\caption{Intensity plot of the multilayer spectral function (\ref{eq:spectral_multi})
in the $\om$ (in units of $\Lambda$) versus in-plane momentum $k_{||}$ (in
units of $\Lambda/\vf$) plane, for different values of the transverse
momentum $k_{\perp}$ (in units of $1/d$) and impurity concentration. Upper row: $\nimp = 10^{-4}$; lower row: $\nimp = 5 \times 10^{-3}$.
}
\label{fig_spectral_multilayer}
\end{figure}

\begin{figure}[htb]
\includegraphics[scale=0.43]{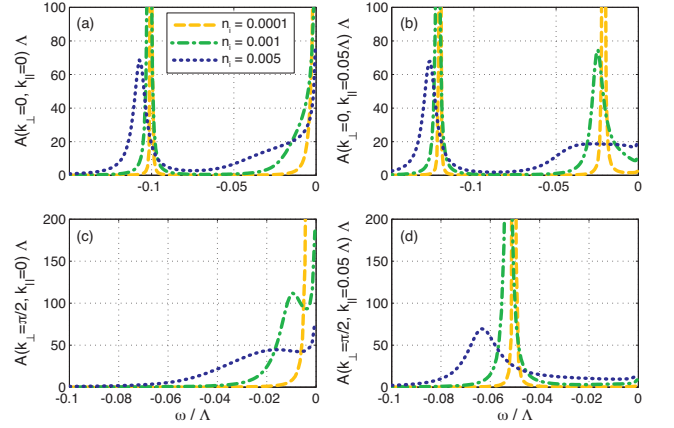}
\caption{(Color online) Constant $\vk$ cuts of the spectral function in 
the multilayer as a function of the frequency (in units of $\Lambda$).
{\bf (a)} $\kpe = 0$, $\kpa=0$, {\bf (b)} $\kpe = 0$, $\kpa=0.05 \Lambda$, 
{\bf (c)} $\kpe = \pi/2$, $\kpa=0$, {\bf (d)} $\kpe = \pi/2$, $\kpa=0.05 
\Lambda$.}
\label{fig_kcuts}
\end{figure}

{\it Infinite stack}. 
The extension of the bilayer model to multilayers is straightforward.
Upon a Fourier transformation in the $c$-direction we can immediately
use the the Hamiltonian in Eq.~(\ref{eq:Hkin0bilayer}) with
${\bf k}=(k_x,k_y,k_{\perp})$, where $k_{\perp}$ is the momentum
along the $c$-axis ($-\pi/2< \kpe d \leq \pi/2$), and make the substitution 
$\tp \rightarrow 2 \tp \cos(\kpe d)$, where $d \approx 2.5 a$ 
is the interlayer distance \cite{Wallace47}. The calculations 
of the DOS and self-energies proceed as previously but with one 
additional $\kpe$-integral \cite{Nilsson2006transport_long_short}.
The electron spectral function, that is measurable in angle resolved
photo-emission (ARPES), is given by:
\begin{equation}
  \label{eq:spectral_multi}
  A(\vk,\om) = -(2/\pi) 
  \Ima \bigl[ \GDA(\vk,\om) + \GDB(\vk,\om) \bigr].
\end{equation}
In Fig.~\ref{fig_spectral_multilayer} we present an intensity plot of
the spectral function for three values of the perpendicular momentum
and two impurity concentrations. In Fig.~\ref{fig_kcuts} we show four 
constant $\vk$-cuts. It is clear that disorder leads to broad peaks 
and that the high-energy branch is less affected by the disorder than the low-energy branch. Electron-electron interactions lead to an 
extra contribution to the self-energy (not included
in the plots) that scales linearly with frequency
\cite{GGV92_short}, $\Ima \Sigma_{ee}(\om) \propto |\om|$. Hence, disorder leads
to a quasiparticle damping that increases at low frequencies, while
the electron-electron contribution increases at high frequencies, 
the final result is a quasiparticle decay rate that has a {\it minimum} at a finite frequency. This highly non-Fermi liquid behavior was also found
in the case of single layer graphene \cite{nuno2006_short} and is
present in {\it all} multilayer systems.
Notice also that the peak positions are shifted by disorder. Another 
interesting feature is the appearance of a new peak in the spectrum near 
$\kpe d = \pi/2$, which is clearly visible in Fig.~\ref{fig_kcuts}(c) 
for $\nimp = 10^{-3}$. This extra peak is due to the fact that the real part 
of the self-energies can act as a ``mass'' term in the Dirac equation 
leading to the formation of a ``pseudogap''. For even higher values of
the disorder the peak goes away and only a broad shoulder remains.
Once again, we can clearly see violations of Fermi liquid theory.


\begin{figure}[htb]
\includegraphics[scale=0.45]{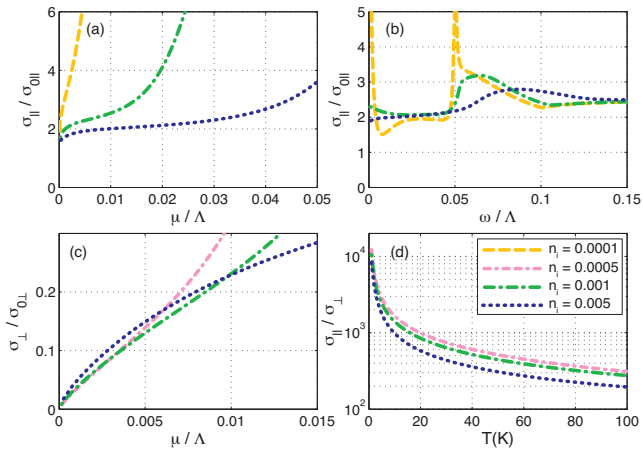}
\caption{(Color online) 
  {\bf (a)} 
  In-plane DC conductivity in the bilayer as a
  function of the chemical potential (in units of $\Lambda$).
  {\bf (b)}
  In-plane conductivity in the bilayer as a function of frequency at
  $T=300\text{K}$ and $\mu = 0$ (in units of $\Lambda$).
  {\bf (c)}
  Perpendicular DC conductivity in the multilayer as a function of the
  chemical potential (in units of $\Lambda$). 
  {\bf (d)} Semi-log plot of the transport
  anisotropy ($\sigma_{||}/\sigma_{\perp}$) 
  in the multilayer as a function of
  the $T$ (in K) at $\mu =0$.
}
\label{fig_conductivities}
\end{figure}

{\it Transport properties}. 
The conductivity can be computed from the Kubo formula:
\begin{equation}
  \label{eq:kubo1}
  \sigma(\om) =\!\!\!\int\!dt  e^{i \om t} \langle [J(t),J(0)]\rangle/(S \omega) \!\!=\!\! i \Pi(\om)/(\om + i\delta),
\end{equation}
where $J$ is appropriate the current operator, and $\Pi(\om)$ the
associated current-current correlation function. The current operators 
are dictated by gauge invariance \cite{kotliar2003} and are computed 
using the Peierls substitution. For instance, for plane 1 with the 
current in the $x$ direction we find:  $J_{x1} = -i \vf e \sum_{\vk} \bigl[
   e^{i \pi /3} c_{A_1,\vk}^{\dag} c_{B_1,\vk}^{\,}
 - e^{- i \pi /3} c_{B_1,\vk}^{\dag} c_{A_1,\vk}^{\,} \bigr]$; 
for the multilayer, the c-axis current operator is: 
$J_{\perp} = -2 e \tp d \sum_{\vk} \sin(k_{\perp}) \bigl[ c_{A_1,\vk }^{\dag} 
c_{A_2,\vk}^{\,} + c_{A_2,\vk }^{\dag} c_{A_1,\vk}^{\,} \bigr]$.
We evaluate the current-current correlators with the disorder-averages 
propagators in the Matsubara formalism \cite{mahan}. For the real part 
of the frequency-dependent conductivity we find:
\begin{equation}
  \label{eq:conductivity2}
  \Rea \bigl[ \sigma(\om) \bigr]/\sigma_0 \!\!=\!\!\! \int\!\!d\e (\nf(\om)-\nf(\om+\e)) \Xi(\e, \e + \om)/\om,
\end{equation}
where $\nf(\e)$ is the Fermi-Dirac function, $\sigma_0$ is the unit of 
conductance  ($\sigma_{0 ||} = 4 e^2/ \pi h$ for the bilayer, and 
$\sigma_{0 ||} = \sigma_{0 \perp} (\vf / 2 \tp d)^2 $ for the multilayer), 
$\Xi$ is a kernel that we have evaluated \cite{Nilsson2006transport_long_short}. Note that since we model the 
impurities as purely local, there are no vertex corrections.
We present some of our results in Fig.~\ref{fig_conductivities}.

The DC conductivities are given by 
$\sigma_{\text{DC}} = \sigma_0 \, \Xi(\mu,\mu)$ where $\mu$ is the
chemical potential. In the bilayer:
\begin{eqnarray}
\Xi_{\text{DC} ||} &=& 
  2 \int_{0}^{\Lambda ^2} d(\vf^2 k^2)  
  \left\{
  \Ima \bigl[ \GDA (\mu,\vk) \bigr]
  \Ima \bigl[ \GDB (\mu,\vk) \bigr] \right.
\nonumber \\
&+& \left.  \Ima \bigl[ \GR_{\text{A}_1 \text{B}_2} (\mu,\vk) \bigr]
  \Ima \bigl[ \GR_{\text{A}_2 \text{B}_1} (\mu,\vk) \bigr]
\right\} \, .
\end{eqnarray}
Upon taking the limit $\mu \rightarrow 0$, using the asymptotic
expressions for the self-energies in Eq.~(\ref{eq:S_smallOm}), 
the off-diagonal propagators drop out and the in-plane DC conductivity 
per plane acquires a universal minimum value given by:
\begin{equation}
  \label{eq:universal_minimum}
  \sigma_{||,{\rm min}} = (3/\pi) (e^2/h),
\end{equation}
independent of the impurity concentration. Hence, as in the case of 
the single layer \cite{nuno2006_short}, we find a universal
conductivity minimum albeit with a different value (in the 
single layer one finds $\sigma_{{\rm min}} = (4/\pi) (e^2/h)$).
This result shows that in these systems the in-plane conductivity is always
of order $e^2/h$ per plane and disorder independent.

The frequency-dependence of the conductivity in the bilayer shows 
some structure. For the cleaner systems a Drude-like peak appears 
at $\om = 0$ due to thermally excited carriers. The second peak at $\om = \tp$
is due to interband transitions. The perpendicular transport in the 
multilayer has the amazing property that close enough to half-filling the 
transport is enhanced by disorder, as can be seen in 
Fig.~\ref{fig_conductivities} (c). Also, as one can clearly see in 
Fig.~\ref{fig_conductivities} (d), the anisotropy ratio becomes very large 
as the Dirac point is approached, and the cleaner the system the larger the 
anisotropy. Using a similar asymptotic expression as in 
Eq.~(\ref{eq:S_smallOm}) one expects that the anisotropy diverges as 
$T^{-2/3}$ exactly at the Dirac point. For small, but finite values of the 
chemical potential, the anisotropy is still enormous but saturates at a 
finite value at $T=0$. Notice, however, that at high temperatures 
electron-phonon scattering (not discussed here) becomes important and 
can substantially modify the transport.

{\it Conclusions}. In conclusion, we have presented results for the
electronic properties of disordered graphene multilayers showing
that the behavior cannot be described by Landau's Fermi liquid
theory of metals. The unconventional behavior includes
divergent self-energies near the Dirac point, the
vanishing density of states on the A sublattice, the non-intuitive
feature that disorder can increase the out-of-plane transport
and the high anisotropy of the system near half-filling.
These properties show that graphene multilayers are a new class
of materials with an unusual metallic state \cite{lanzara_new}.

\begin{acknowledgments}
We thank A.~Lanzara for sharing ref.~[\onlinecite{lanzara_new}] prior
to its publication.
A.H.C.N. is supported through NSF grant DMR-0343790.
N.M.R.P. thanks ESF Science Programme INSTANS 2005-2010
and FCT under the grant POCTI/FIS/58133/2004. F.G. acknowledges
funding from MEC (Spain) through grant FIS2005-05478-C02-01, and the European Union contract 12881 (NEST).
\end{acknowledgments}

\bibliography{graphite12.bib}

\end{document}